\documentclass[12pt]{article}
\author{Author Name}
\DeclareUnicodeCharacter{03A6}
{\ensuremath{\Phi}}

\usepackage{amsmath}

\usepackage{amssymb}

\usepackage{doi}

\usepackage{longtable}

\pdfoutput=1
\usepackage{graphicx}

\usepackage{float} 

\usepackage{cite}
\usepackage{tabularx}

\usepackage{microtype} 

\usepackage[utf8]{inputenc}
\usepackage{longtable}
\usepackage{array}
\usepackage{ragged2e}
\usepackage[utf8]{inputenc}
\usepackage{enumitem}
\usepackage{xurl}

\usepackage{float}

\title{Trusted Identities for AI Agents: Leveraging Telco-Hosted eSIM Infrastructure}
\author{Sebastian Barros}
\date{April 13th , 2025}  

\begin{document}
\renewcommand{\arraystretch}{1.2}

\maketitle

\begin{abstract}
The rise of autonomous AI agents in enterprise and industrial environments introduces a critical challenge: how to securely assign, verify, and manage their identities across distributed systems. Existing identity frameworks based on API keys, certificates, or application-layer credentials, lack the infrastructure-grade trust, lifecycle control, and interoperability needed to manage agents operating independently in sensitive contexts.

In this paper, we propose a conceptual architecture that leverages telecom-grade eSIM infrastructure, specifically hosted by mobile network operators (MNOs), to serve as a root of trust for AI agents. Rather than embedding SIM credentials in hardware devices, we envision a model where telcos host secure, certified hardware modules (eUICC or HSM) that store and manage agent-specific eSIM profiles. Agents authenticate remotely via cryptographic APIs or identity gateways, enabling scalable and auditable access to enterprise networks and services.

We explore use cases such as onboarding enterprise automation agents, securing AI-driven financial systems, and enabling trust in inter-agent communications. We identify current limitations in GSMA and 3GPP standards, particularly the device-centric assumptions—and propose extensions to support non-physical, software-based agents within trusted execution environments. This paper is intended as a conceptual framework to open discussion around standardization, security architecture, and the role of telecom infrastructure in the evolving agent economy.
\end{abstract}

\section{Introduction}

Autonomous AI agents are rapidly emerging as integral components of enterprise and industrial workflows. These agents-software entities capable of decision-making and goal-directed actions, are now used for automating customer support, optimizing logistics, analyzing financial markets, and managing IT infrastructure \cite{lecun2022path}. Unlike traditional software systems, autonomous agents often operate with a degree of independence from human oversight, acting as semi-autonomous actors within digital environments.

As these agents proliferate, the need for secure, verifiable, and manageable identities becomes critical. Identity is the foundation of access control, auditing, and trust in distributed systems. Yet, most existing identity frameworks were designed for human users (e.g., username/password, OAuth), physical devices (e.g., X.509 certificates, TPMs), or static service accounts within tightly controlled application domains. These mechanisms lack the flexibility, life cycle management, and infrastructure-grade assurance needed to support dynamic, distributed agent populations operating across organizational and cloud boundaries.

In parallel, the telecommunications industry has spent decades solving a related problem: securely identifying and managing billions of non-human endpoints such as mobile devices, SIMs, and more recently, IoT nodes using standardized, tamper-resistant, and remotely provisioned identity modules. The emergence of embedded SIM (eSIM) and its extensions for IoT, such as GSMA SGP.32, offer a scalable, remotely manageable, cryptographically secure identity layer already deployed globally \cite{gsma_sgp32, esim_overview}. These technologies provide secure on boarding, life cycle control, and interoperability for non-human endpoints at massive scale.

This paper proposes a conceptual architecture that adapts telecom-grade eSIM infrastructure to the identity needs of virtual AI agents. In our model, mobile network operators (MNOs) host certified secure hardware such as embedded UICC modules or hardware security modules (HSMs), in their data centers. Rather than embedding SIM credentials in physical devices, unique eSIM profiles are assigned to autonomous agents, which authenticate remotely via secure protocols. The result is an infrastructure that provides scalable, high-assurance identity as a service-backed by the same root of trust used to secure the global mobile ecosystem.

Our contribution is a technically grounded but conceptual framework that explores how this approach could be implemented, identifies current limitations in GSMA and 3GPP standards, and proposes specific extensions to support non-physical, software-based agents. We present practical use cases in enterprise automation, financial services, and industrial applications, and analyze the security implications of assigning telecom-managed identities to software entities.

The remainder of this paper is structured as follows. We begin by reviewing the current capabilities and limitations of eSIM technology and identity management frameworks, particularly in the context of non-human entities. We then introduce our proposed architecture, in which telco-hosted secure hardware modules manage agent-specific eSIM profiles and enable remote authentication for autonomous software agents. This is followed by a discussion of practical enterprise and industrial use cases where such a framework could be deployed. We then analyze the security implications and trust model associated with using telecom infrastructure to authenticate software agents. The paper continues with an assessment of current standardization gaps within GSMA and 3GPP specifications and offers concrete recommendations for extending these standards. We conclude by comparing our proposal with existing approaches to agent identity and discussing future research directions.

\section{Background}

\subsection{The Telecom SIM Infrastructure: From Physical Devices to eSIM}

The Subscriber Identity Module (SIM) has long served as the foundational element for identity and authentication in mobile telecommunications. A SIM securely stores a user’s International Mobile Subscriber Identity (IMSI) and a secret authentication key (Ki), which are used in a challenge-response protocol to authenticate the device to the mobile network. This architecture has provided scalable, carrier-grade identity infrastructure for billions of mobile users worldwide \cite{3gpp_33501}.

Over the past decade, SIM technology has evolved significantly. The traditional removable SIM card has been replaced in many devices by the embedded SIM (eSIM), a programmable chip soldered onto a device’s motherboard. Unlike physical SIM cards, eSIMs are remotely provisioned using over-the-air protocols defined by the GSM Association (GSMA). The provisioning architecture includes the Subscription Manager Data Preparation (SM-DP+), which prepares encrypted operator profiles, and the Subscription Manager Secure Routing (SM-SR), which securely delivers and activates profiles on the eUICC (embedded Universal Integrated Circuit Card) \cite{gsma_sgp32}.

eSIM functionality is further extended by the integrated SIM (iSIM), which is embedded directly within the system-on-chip (SoC) of a device, reducing size and cost for constrained environments such as wearables and industrial IoT. All three SIM models, removable SIM, eSIM, and iSIM adhere to the core principle of providing a secure, tamper-resistant environment for storing identity credentials and running cryptographic functions.

The use of eSIMs is now mainstream in consumer devices such as smartphones, tablets, and smartwatches. Apple, for example, fully transitioned to eSIM-only iPhones in the U.S. market with the iPhone 14 series in 2022 \cite{apple_esim2022}. In parallel, eSIM adoption has accelerated across industrial and IoT sectors, where device deployment at scale and the need for remote management make eSIM provisioning especially valuable. The GSMA’s latest standard, SGP.32, introduces a lightweight architecture specifically designed for large-scale IoT device onboarding, eliminating reliance on SMS and enabling bulk provisioning over constrained networks \cite{gsma_sgp32}.

At its core, the SIM infrastructure offers:
\begin{itemize}
  \item Cryptographically secured identity and mutual authentication with the mobile network
  \item Secure storage of long-term secrets (e.g., Ki)
  \item Remote lifecycle management (provisioning, activation, revocation)
  \item Interoperability across mobile operators via GSMA root-of-trust certification
\end{itemize}

These properties make eSIM one of the most battle-tested and scalable identity systems in the world. However, as we explore in later sections, current standards remain tightly bound to physical devices and do not account for purely virtual entities such as autonomous AI agents.

\subsection{Identity Management with eSIM: Strengths and Industry Applications}

eSIM technology is more than a connectivity enabler; it is a cryptographically grounded identity mechanism backed by one of the most mature and globally trusted infrastructure layers in digital systems. At the center of this trust model is the concept of mutual authentication: a device proves possession of a secret key (Ki), provisioned and managed by the operator, and in turn receives authentication from the mobile network using the Authentication and Key Agreement (AKA) protocol \cite{3gpp_33501}.

In this architecture, the SIM serves as both a secure credential store and a hardware-enforced execution environment. The cryptographic functions required to authenticate with the network, such as generating responses to challenge values—are executed entirely within the protected SIM hardware, ensuring that the secret key never leaves the secure boundary.

Enterprises and operators have extended this model into the IoT domain. For example, an eSIM-enabled industrial sensor deployed in a factory can be remotely provisioned with a local mobile network profile, securely identified using its IMSI, and granted role-based access to specific data streams or control systems. The GSMA's SGP.32 standard further simplifies this process by enabling large-scale, zero-touch provisioning of IoT devices over lightweight protocols that do not require SMS or user interfaces \cite{gsma_sgp32}.

This trust model has also been adopted beyond cellular connectivity. In frameworks like IoT SAFE (SIM Applet For Secure End-to-End Communication), the SIM serves as a root of trust for cloud applications: it can generate TLS client certificates, encrypt telemetry data, and authenticate to cloud IoT platforms using keys stored in the eUICC. Telecom identity is thus being repurposed to establish secure, verifiable interactions between devices and services across both private and public networks.

These features such as hardware-protected identity, cryptographic integrity, remote manageability, and global interoperability make eSIM an appealing substrate for scalable identity systems. However, as powerful as this model is, it is fundamentally tied to physical devices, which presents significant challenges when considering the identity of non-physical, cloud-native AI agents.

\subsection{Current Limitations for Autonomous AI Agents}

While the eSIM infrastructure has matured around device-based deployments, its underlying assumptions create friction when applied to software agents. The core limitation is that eSIMs are bound to hardware: each profile is linked to a specific physical secure element, such as a soldered eUICC or integrated iSIM, identified by a unique EID (Embedded Identity). This binding is foundational to the security model; credentials are protected by the physical boundary of the SIM hardware.

Autonomous software agents, by contrast, are not tied to a fixed device. They are virtual processes that can be instantiated, migrated, or terminated dynamically across cloud environments. An agent might exist for milliseconds as part of a serverless function, or persist for months as a continuously learning system deployed across regions. These agents do not have EIDs or ICCIDs; they exist only within abstract computational environments that lack the certified, tamper-resistant properties expected by SIM provisioning systems.

The second limitation is lifecycle coupling. eSIM provisioning flows (e.g., GSMA SM-DP+ architecture) assume relatively stable device identities. Devices are onboarded once, managed for months or years, and decommissioned predictably. Autonomous agents, especially those used in AI pipelines, may be highly transient. Requiring a traditional provisioning flow for each ephemeral instance is operationally infeasible.

Security presents a third and more subtle challenge. Without hardware anchoring, agent credentials become software-stored secrets vulnerable to theft, replication, or impersonation. Even if an agent were issued an IMSI and key in software, the lack of a secure execution boundary means the credential could be exfiltrated or misused. This breaks the core principle of telecom-grade trust, where possession of a credential implies physical integrity.

Finally, current GSMA and 3GPP specifications do not account for virtualized identities. There is no formal support for:
\begin{itemize}
  \item Assigning SIM profiles to non-physical entities
  \item Provisioning identity across execution environments (e.g., virtual machines, containers)
  \item Multi-tenancy or shared secure element models
  \item APIs for agent-to-network authentication without radio hardware
\end{itemize}

In short, while eSIMs offer one of the most secure and scalable identity models in existence, their applicability to virtual AI agents is blocked by assumptions of hardware-bound identity, static provisioning flows, and device-centric trust semantics. Overcoming these limitations requires a rethinking of how telecom infrastructure can act as a trust anchor for dynamic, software-native actors.

\begin{table}[ht]
\centering
\caption{Comparison of eSIM Identity for Devices vs Autonomous AI Agents}
\label{tab:comparison}
\begin{tabularx}{\textwidth}{|l|X|X|}
\hline
\textbf{Property} & \textbf{eSIM for Devices} & \textbf{Target for AI Agents} \\
\hline
Identity Binding & Bound to physical secure element (eUICC/iSIM) & Decoupled from hardware; bound to virtualized runtime \\
\hline
Provisioning Model & Static, long-term profile activation & Dynamic, on-demand provisioning tied to agent lifecycle \\
\hline
Security Boundary & Tamper-resistant hardware module & Trusted execution environment or verified enclave \\
\hline
Lifecycle Duration & Persistent for months or years & Short-lived or elastic (seconds to weeks) \\
\hline
Execution Context & One profile per device, hardware-bound & Multi-profile, multi-agent on shared infrastructure \\
\hline
Standard Support & Fully covered by GSMA (e.g. SGP.32) & No existing standard for virtual or software agents \\
\hline
Network Interface & Assumes physical radio/modem & Purely IP-based; agent may not access radio at all \\
\hline
\end{tabularx}
\end{table}

\section{Proposed Architecture: Hosted eSIM Identity for AI Agents}

\subsection{Decoupling Identity from Devices: Conceptual Overview}

To adapt eSIM infrastructure for autonomous AI agents, we must address three foundational challenges: (1) the absence of physical secure elements in virtual environments, (2) the lack of standards for non-device-based provisioning, and (3) the inability of agents to natively perform SIM-based authentication without radio hardware.

\textbf{First}, all GSMA and 3GPP eSIM profiles are currently provisioned to a certified hardware element—typically a soldered eUICC or integrated iSIM present on a physical device \cite{gsma_sgp32, 3gpp_33501}. These secure elements are tamper-resistant and host the cryptographic functions needed for authentication with the mobile core. Virtual AI agents, however, lack such hardware by default. They execute in software-defined environments such as virtual machines, containers, or serverless runtimes, none of which include a certified eUICC.

\textbf{Second}, existing eSIM provisioning flows (e.g., SM-DP+ and SM-SR architecture) require a unique hardware identifier (EID) to establish a secure binding between the subscription and a device \cite{esim_overview}. In the case of software agents, there is no physical EID to anchor the profile. Moreover, these provisioning workflows are optimized for stable, long-lived devices and not ephemeral agents whose lifespans may be measured in seconds or minutes.

\textbf{Third}, the authentication protocol used by SIMs; typically based on the Authentication and Key Agreement (AKA) protocol defined in 3GPP TS 33.501—assumes the presence of a radio interface (e.g., LTE or 5G) and a modem stack capable of issuing authentication requests to the mobile core \cite{3gpp_33501}. AI agents may not have any radio access at all, nor a modem stack, rendering traditional SIM-based authentication workflows inapplicable.

\bigskip

To overcome these constraints, we propose a decoupled architecture where the eSIM is not embedded in the agent's execution environment but instead hosted externally by a trusted telecom provider. This allows us to preserve the security properties of SIM-based authentication without requiring the agent to possess or emulate physical SIM hardware.

In this model, the Mobile Network Operator provisions SIM profiles to certified secure elements located in its own infrastructure—either physical eUICCs or FIPS 140-2 compliant Hardware Security Modules (HSMs) with embedded UICC applets. These secure modules are isolated, audit-logged, and managed in accordance with GSMA security standards.

Each AI agent is assigned a unique profile and interacts with the SIM via a secure identity gateway. This gateway exposes a set of controlled cryptographic operations (e.g., generating AKA responses, signing nonces) through an authenticated API. Crucially, the agent does not have direct access to the key material (Ki); it only delegates identity-related operations to the telecom-hosted module. This mirrors the design of Trusted Platform Modules (TPMs) in secure computing: the trust lies not in ownership of the key, but in the integrity of the cryptographic boundary that controls it.

This architecture introduces a new abstraction layer: the \textit{virtual identity binding}. Instead of associating a SIM profile with a physical device, the identity is bound to a cryptographically verifiable software context—e.g., an agent running inside a verified enclave, or a container signed and attested by an enterprise control plane. A Telco-issued SIM profile becomes an identity "token" that can be invoked through secured APIs, not through modem commands.

\begin{figure}[H]
\centering
\includegraphics[width=0.75\linewidth]{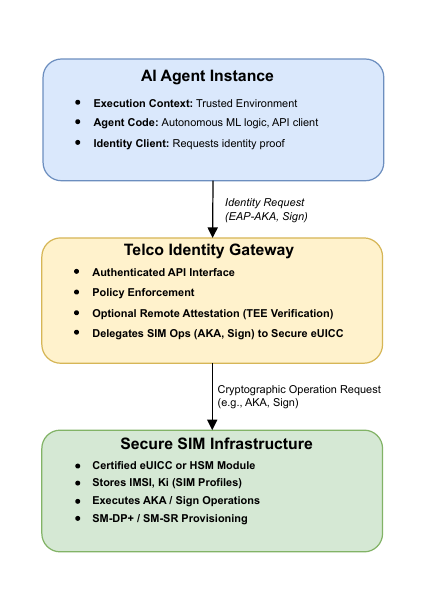}
\caption{Conceptual model: virtual AI agent authenticating via telco-hosted eSIM infrastructure}
\label{fig:architecture}
\end{figure}

In this configuration, identity becomes a service: the SIM remains telecom-issued, cryptographically secure, and remotely managed—but decoupled from physical instantiation. The agent authenticates using telecom infrastructure without ever physically possessing the SIM.

This model retains the strongest properties of traditional eSIM:
\begin{itemize}
  \item \textbf{Hardware-rooted trust:} all cryptographic operations are anchored in certified hardware hosted by the operator.
  \item \textbf{Provisioning discipline:} agent onboarding and deactivation are managed through GSMA-compliant remote lifecycle controls.
  \item \textbf{Interoperability:} the identities are valid across operator networks, leveraging existing roaming and trust frameworks.
\end{itemize}

This decoupling of assigning telecom-grade identity to software agents via hosted infrastructure, forms the core architectural premise of this work. In the next subsection, we explore how this identity can be provisioned, authenticated, and securely bound to agent runtimes.

\subsection{Secure Profile Management and Remote Agent Authentication}

In a traditional eSIM deployment, the SIM profile containing the International Mobile Subscriber Identity (IMSI), the cryptographic key (Ki), and operator credentials is provisioned directly onto a device-bound secure element, such as an eUICC or iSIM. In our architecture, these SIM profiles are instead installed on telco-hosted secure elements (e.g., certified HSMs or eUICCs within a data center), and are accessed by autonomous agents over a secure interface. This redefines the SIM not as a physical object, but as a remotely managed identity service.

\textbf{Provisioning.} An enterprise wishing to onboard an autonomous agent requests a SIM profile for it via the Telco's provisioning interface. This is conceptually similar to device enrollment workflows in GSMA’s SGP.32 standard \cite{gsma_sgp32}, but decoupled from hardware identifiers like EIDs. Instead, the agent may be identified by a public key, attested environment ID (e.g., TPM or enclave attestation), or a signed deployment manifest. Once verified, the Telco issues an IMSI and Ki, and stores this profile securely in its hosted secure element infrastructure. This step may also register metadata about the agent’s expected execution boundary (e.g., TEE ID, container fingerprint, or enterprise namespace).

\textbf{Authentication.} When the agent needs to authenticate (e.g., to a network, enterprise service, or peer agent), it issues an API request to the Telco Identity Gateway. The request includes:
\begin{itemize}
  \item A signed challenge from the remote service (e.g., EAP-AKA nonce or JWT header)
  \item A reference to the agent’s SIM identity (e.g., IMSI or profile ID)
  \item Optionally, an attestation token proving the agent is running inside a verified context
\end{itemize}

Upon receiving the request, the Identity Gateway verifies access control rules (rate limits, validity, time window), and delegates the request to the secure module storing the SIM profile. The SIM applet within the secure module computes the appropriate cryptographic response, such as an AKA challenge-response pair or a digital signature, without exposing the key material to the agent or application layer.

This model preserves the most important cryptographic property of the SIM: the Ki key is never exported or accessible, not even to the agent that uses it. Instead, the agent acts as a client of a cryptographic signing service, with hardware-backed operations executed inside telecom-grade secure infrastructure.

This interaction can be modeled as a constrained identity API, with endpoints such as:

\begin{itemize}
  \item \texttt{POST /identity/sign} – returns a signed challenge or nonce
  \item \texttt{POST /identity/authenticate} – executes a complete AKA challenge flow
  \item \texttt{GET /identity/status} – verifies whether an identity is currently active and bound
\end{itemize}

\textbf{Response flow.} After the secure element produces the signed output, it is relayed back to the agent through the Identity Gateway. The agent can then forward this to the relying service or authentication server, just as a device would respond in a native SIM-based authentication flow. This enables the agent to prove its identity without ever touching the key or the SIM profile directly.

\textbf{Security benefits.} By preserving the separation between execution (agent) and identity (hosted SIM), this architecture prevents key leakage, supports revocation, and enables multi-party auditability. It aligns with modern confidential computing practices, where keys reside in HSMs or enclaves and are only used through API calls—not loaded into application memory.

The result is an end-to-end trusted identity path: from GSMA-compliant profile provisioning, to cryptographically attested use by autonomous software agents without ever exposing physical keys outside of certified telecom infrastructure.

\subsection{Trust Anchors, Execution Boundaries, and Delegation Policies}

To ensure the integrity of identity-based operations executed on behalf of an AI agent, it is not sufficient to control cryptographic access alone. One must also ensure that the agent invoking the Telco Identity Gateway is operating within a verifiable and constrained execution boundary. This subsection explores how trust anchors are established, how execution contexts are validated, and how delegation policies restrict access to telecom-grade identities.

\textbf{Binding Identity to Execution Context.} In traditional device-based systems, identity is anchored to physical hardware—e.g., the SIM is embedded in a unique, tamper-resistant chip. In our architecture, we achieve a similar anchoring effect by requiring the AI agent to run within a verifiable environment, such as a Trusted Execution Environment (TEE), container with measured boot, or hardware-isolated enclave.

Before granting access to SIM profile operations, the Telco Identity Gateway verifies a signed attestation token produced by the agent's runtime. This token may include:
\begin{itemize}
  \item A cryptographic measurement of the code running (e.g., enclave hash)
  \item Environment ID or VM UUID
  \item Validity window and timestamp
  \item Signature from a hardware root-of-trust (e.g., Intel SGX, ARM TrustZone, AWS Nitro) 
\end{itemize}

Attestation protocols such as Intel’s EPID or DCAP, or AWS Nitro attestations over KMS-backed trust chains, provide verifiable proof that the agent operates within a known, untampered, and isolated boundary. These attestation claims are checked by the Identity Gateway before allowing any cryptographic operation to proceed.

\textbf{Delegation Policies.} Once identity has been bound to a runtime, Telcos can define fine-grained delegation policies that govern how, when, and for what purpose an identity can be used. Policies may include:

\begin{itemize}
  \item \textit{Rate Limits:} Max number of cryptographic operations per agent per time unit.
  \item \textit{Time-bound Access:} Profiles expire or rotate after a given lifetime.
  \item \textit{Role-based Constraints:} E.g., “This profile can only be used for signing inter-agent messages, not accessing telecom core.”
  \item \textit{Location-based or IP-scoped Access:} Allow usage only from certain geographic or network locations.
  \item \textit{Execution Attestation Constraints:} Only allow usage when verified TEE measurements match a signed list.
\end{itemize}

These policies are enforced at the Telco Identity Gateway level, and are backed by signed profile metadata managed during provisioning. Profile status and delegation parameters can be queried and updated via secure Telco APIs, which act as an administrative overlay atop the SIM infrastructure.

\textbf{Auditability and Revocation.} Because the agent interacts with SIM credentials only via remote API calls, all identity usage can be logged and monitored centrally. This enables full auditability critical for compliance in sectors like finance, defense, and regulated industries. If an agent is compromised, its ability to use its assigned identity can be revoked immediately by deactivating the profile or denying future access through the gateway. No changes to the agent’s internal code or environment are required.

\textbf{Security Posture.} This architecture extends the traditional telco model, where physical possession equates to identity ownership, into a software-native paradigm where attestation replaces possession. The SIM profile becomes usable only within a cryptographically verified runtime, under enforceable policy conditions. This model aligns with modern principles of zero trust and

\subsection{Telco Deployment Models and Enterprise Integration}

The architectural model described in this paper is flexible and can be deployed in multiple configurations depending on regulatory, latency, or operational considerations. In all cases, the Telco remains the custodian of the identity infrastructure, while enterprises or developers act as consumers of the identity-as-a-service interface. We outline three representative deployment models and describe their implications.

\textbf{Model 1: Telco-Hosted Centralized Identity Service.} In the most straightforward deployment, the Telco provisions and stores all eSIM profiles in its own secure infrastructure e.g., HSM clusters or datacenter-hosted eUICCs. The Identity Gateway is exposed as a multi-tenant API endpoint over a secure channel (e.g., mTLS or VPN). This model mirrors existing cloud-hosted SIM platforms used in IoT and M2M, where devices authenticate using telecom infrastructure without owning a physical SIM \cite{gsma_sgp32}.

This model is ideal for early-stage deployments, multi-tenant environments, or cross-enterprise scenarios where operational control rests with the Telco. It offers economies of scale, centralized logging, and the ability to reuse existing SIM provisioning infrastructure.

\textbf{Model 2: Enterprise-Edge Deployment.} For latency-sensitive or compliance-bound use cases, the Telco may distribute SIM infrastructure closer to the enterprise. This can take the form of Telco-certified edge nodes (e.g., a 1U rack-mounted eUICC server) hosted inside an enterprise datacenter or at the customer edge. Profiles are still provisioned and managed by the Telco but physically reside within enterprise-controlled hardware.

This model is attractive for on-premise environments with strict data sovereignty rules, such as defense, healthcare, or industrial automation. It also enables ultra-low-latency authentication for agents operating on edge compute nodes.

\textbf{Model 3: Federated Trust with Cloud Providers.} In a more advanced configuration, the Telco Identity Gateway federates trust with public cloud providers. Rather than directly hosting the SIM profile, the Telco may authorize operations based on attested enclave environments (e.g., AWS Nitro Enclaves, Azure Confidential Compute). The SIM cryptographic key remains inside Telco custody, but a federated attestation scheme allows remote use of the identity by cloud-based agents.

This model mirrors federated identity systems such as OAuth2 and SAML but applies them to SIM-backed identity infrastructure. It allows AI agents running in cloud-native environments to access Telco-managed credentials without moving the key, only the proof-of-execution environment.

\textbf{Enterprise Integration.} Regardless of deployment model, enterprise systems integrate with the identity service via standard APIs. These APIs may be embedded in agent SDKs or exposed through enterprise IAM systems such as API gateways, service meshes, or Kubernetes admission controllers. The use of signed attestation tokens as proof-of-execution enables compatibility with zero trust policies and workload-level access control.

Integration with existing enterprise systems may include:
\begin{itemize}
  \item \textit{Service mesh plug-ins:} Inject SIM-backed identity into network requests.
  \item \textit{CI/CD hooks:} Automatically provision identities for agent workloads at deployment time.
  \item \textit{Policy engines:} Enforce that only signed workloads with valid attestation may trigger SIM operations.
\end{itemize}

This architecture aligns with trends in confidential computing and programmable telecom networks. It creates a pathway for Telcos to become providers of decentralized, hardware-rooted identity for agents, with a monetization layer built on SLAs, usage tiers, and authentication volume.

\section{Use Cases and Enterprise Applications}

The proposed architecture enables a wide range of secure, infrastructure-grade identity services for autonomous agents across verticals. By decoupling SIM credentials from physical devices and anchoring identity in telecom-hosted infrastructure, this model opens new possibilities for integrating AI agents into enterprise and industrial systems. We describe three representative scenarios where this architecture delivers unique value.

\subsection{Enterprise Automation Agents with SIM-Grade Identity}

Large enterprises are increasingly deploying AI agents to autonomously manage IT infrastructure, customer support workflows, document routing, and internal ticket resolution. In many cases, these agents operate across hybrid cloud environments and interact with sensitive systems.

Assigning a telecom-grade identity to these agents allows:
\begin{itemize}
    \item Auditable authentication to APIs and internal services using strong, non-replicable credentials.
    \item Lifecycle-managed access tied to verified execution environments (e.g., verified containers in CI/CD).
    \item Federated identity federation across subsidiaries, clouds, or regulatory domains.
\end{itemize}

For example, an AI agent that monitors enterprise VPN access and triggers alerts could be given an eSIM-backed identity with policy-based usage (e.g., “Can only sign alerts once per hour, and only from this environment”).

\subsection{Secure AI-Driven Financial Systems}

In fintech and banking, AI agents are increasingly used for tasks such as automated credit scoring, fraud detection, transaction monitoring, and regulatory reporting. These agents must meet stringent compliance requirements around authentication, traceability, and non-repudiation.

Hosted eSIM identities enable:
\begin{itemize}
    \item Strong identity proofs that are legally auditable and cryptographically verifiable.
    \item Isolation of agent credentials from the application codebase (zero key exposure).
    \item Instant revocation of identity access upon policy violation or environment compromise.
\end{itemize}

An example scenario involves a credit decisioning agent that interacts with core banking APIs. A telecom-hosted SIM identity ensures that every decision is traceable to a cryptographically bound agent instance operating under an enterprise SLA.

\subsection{Industrial Edge AI and Robotics Integration}

Manufacturing, logistics, and energy sectors increasingly rely on distributed edge AI agents to operate factory equipment, inspect infrastructure, or coordinate autonomous vehicles. These agents are often deployed on shared or transient hardware with limited physical security controls.

This architecture enables:
\begin{itemize}
    \item Secure authentication to edge gateways and control networks from remote agents.
    \item Tamper-evident binding between agent workload and identity (e.g., signed enclave attestation).
    \item Identity persistence and portability across hardware resets or software migrations.
\end{itemize}

For instance, a visual inspection agent deployed on a mobile robot in a mining site could authenticate securely to a central control system, even if re-instantiated frequently or across nodes—without reissuing credentials.

\subsection{Inter-Agent Communication with Verifiable Trust}

As AI agents begin to interact with each other, negotiating, coordinating, or executing transactions establishing trust between them becomes essential. Existing methods (e.g., self-signed certificates or ephemeral keys) lack infrastructure-grade assurance.

By using SIM-backed identities:
\begin{itemize}
    \item Each agent can cryptographically prove its origin and environment.
    \item Communication between agents can be signed, verified, and audited.
    \item Trust relationships can be enforced based on SIM issuer (e.g., telco-level federation).
\end{itemize}

This capability lays the foundation for regulated agent ecosystems—e.g., agent-to-agent API marketplaces, verified supply chain flows, or autonomous service brokers.

\subsection{Expanded Use Case Table: Non-Exhaustive Applications}

While the previous sections focused on flagship enterprise and industrial scenarios, the applicability of SIM-grade identity for software agents extends far beyond. Table~\ref{tab:usecases} presents a non-exhaustive list of possible use cases, categorized by vertical, highlighting how telecom-hosted secure identity could provide unique value in each domain.

\begin{table}[H]
\centering
\caption{Illustrative Use Cases for SIM-Based Identity in AI Agents}
\label{tab:usecases}
\renewcommand{\arraystretch}{1.1}
\small
\begin{tabularx}{\textwidth}{|l|X|X|}
\hline
\textbf{Sector} & \textbf{Agent Role} & \textbf{SIM Identity Benefit} \\
\hline
Healthcare & Triage bots in telemedicine & Auditable access to patient data \\
\hline
Cybersecurity & Threat detection agents & Tamper-proof operational integrity \\
\hline
Smart Cities & Traffic flow optimizers & Signed, trusted inter-agent messages \\
\hline
Insurance & Claim handlers & Time-bound identity with revocation \\
\hline
Retail & Pricing bots & Fair market behavior enforcement \\
\hline
Supply Chain & Customs logistics agents & Federated cross-border trust \\
\hline
Legal Tech & Contract analyzers & Role-based execution guarantees \\
\hline
Cloud Ops & Infra management agents & Execution-bound access and auditing \\
\hline
Content Platforms & Moderation bots & Central policy and access control \\
\hline
Industrial IoT & Digital twin agents & Non-physical but trusted identity \\
\hline
\end{tabularx}
\end{table}

This compact table illustrates the horizontal applicability of agent SIM identity across sectors. Each case reflects environments where strong, telecom-managed credentials offer advantages beyond traditional identity models.

This table is not meant to be exhaustive, but to illustrate the breadth of opportunities where AI agents require identities that are verifiable, policy-enforced, and cryptographically anchored. In many of these cases, existing identity frameworks (API keys, certificates) fall short in terms of revocation, lifecycle management, or regulatory trustworthiness. Telecom-grade SIM identity adapted to the agent economy can fill this gap.

\section{Security and Trust Model}

Security is the foundation of any identity system. The architecture proposed in this paper aims to offer a security posture comparable to traditional SIM-based authentication, while adapting to the unique challenges of virtual AI agents operating in software-defined environments.

\subsection{Threat Model and Assumptions}

This system assumes that attackers may:
\begin{itemize}
    \item Gain access to the runtime environment of the agent
    \item Attempt to spoof, replicate, or replay identity operations
    \item Try to extract or misuse SIM credentials
    \item Compromise the network channel between agent and Telco Identity Gateway
\end{itemize}

To defend against these threats, we assume:
\begin{itemize}
    \item The Telco Identity Gateway and hosted SIM modules are operated within hardened environments (e.g., FIPS 140-2 HSMs, eUICC datacenter appliances)
    \item All API communication between agent and gateway is encrypted and mutually authenticated (e.g., mTLS)
    \item Agent environments support verifiable attestation (e.g., via TEE, VMM signatures, or cloud-native attestation chains)
\end{itemize}

\subsection{Key Isolation and SIM Confidentiality}

A key strength of this model is that SIM credentials, particularly the Ki and IMSI, are never exposed to the agent. All cryptographic operations occur inside a tamper-resistant module under Telco control. This mitigates:
\begin{itemize}
    \item Credential exfiltration via memory dumps or disk access
    \item Agent impersonation or key theft via compromised containers
    \item Replay of past authentication material
\end{itemize}

Because access to these operations is API-mediated and tightly policy-controlled, the risk of unauthorized use is substantially reduced.

\subsection{Execution Integrity via Remote Attestation}

To prevent malicious code from hijacking SIM identity use, this system requires runtime attestation from the agent's execution environment. Depending on deployment model, this may take the form of:
\begin{itemize}
    \item Intel SGX or TDX enclave attestation
    \item AWS Nitro enclave signatures
    \item ARM TrustZone hardware reports
    \item Confidential VM attestation (e.g., AMD SEV, Azure confidential compute)
\end{itemize}

Only if the agent proves it is executing inside a trusted and pre-approved environment will the Identity Gateway authorize SIM operations. This shifts the trust model from physical possession to verifiable software integrity.

\subsection{Policy Enforcement and Rate Limiting}

Every identity request is subject to programmable constraints. These may include:
\begin{itemize}
    \item Per-agent rate limits (e.g., 10 auths/minute)
    \item Time-based validity (e.g., credentials usable for 24 hours)
    \item Role-based limitations (e.g., can only be used for signing, not authentication)
    \item IP/geolocation binding
\end{itemize}

This makes credential use both contextual and enforceable in real time. It also reduces exposure in the event of a partial compromise.

\subsection{Auditability and Revocation}

Because all use of SIM credentials flows through the Identity Gateway, every operation is logged. This enables:
\begin{itemize}
    \item Time-stamped, non-repudiable audit trails
    \item Forensic analysis in case of agent compromise
    \item Immediate revocation by disabling SIM profile access
\end{itemize}

This auditability provides strong support for compliance with regulatory frameworks such as GDPR, HIPAA, and financial risk management standards.

\subsection{Comparison to TPMs, API Keys, and X.509 Certificates}

Compared to local TPMs or application-held X.509 certificates, this architecture offers:
\begin{itemize}
    \item No key exposure to the application layer
    \item Stronger assurance of credential custody (via Telco infrastructure)
    \item Centralized revocation, usage tracking, and SLA-backed availability
\end{itemize}

Unlike API keys or bearer tokens, which can be trivially copied, SIM-based identity operations require verified access to secure modules. This positions Telcos as trust brokers for digital agent identity at infrastructure scale.

\section{Standards Landscape and Comparison to Alternatives}

While the telecom industry has developed robust identity frameworks for physical devices, there is currently no standardized support for provisioning and managing SIM identities for non-physical, software-based agents. The architecture proposed in this paper reveals critical gaps in existing specifications and highlights a new opportunity for Telcos to support the emerging agent economy.

\subsection{Gaps in GSMA and 3GPP Specifications}

GSMA’s eSIM provisioning architecture, including SGP.22 (for consumer devices) and SGP.32 (for IoT), is built on the assumption of device-bound identities. The provisioning flows depend on the presence of an EID (Embedded Universal Integrated Circuit Card Identifier), which is tied to a tamper-resistant hardware module located within a device \cite{gsma_sgp32}. There is currently no provision for assigning SIM profiles to cloud-based, containerized, or serverless agents that lack such physical hardware.

Similarly, 3GPP’s authentication framework (e.g., TS 33.501) is centered around device-initiated access via a radio modem. SIM-based authentication assumes the presence of NAS-layer signaling between the device and the core network \cite{3gpp_33501}. Virtual agents communicating over IP networks have no access to such radio-layer protocols, and cannot natively initiate AKA-based flows.

There are also no standards for:
\begin{itemize}
    \item Remotely attested use of SIM credentials by cloud agents
    \item Delegation of SIM operations via policy-controlled APIs
    \item Hosting of multi-tenant SIM vaults in data centers
    \item Binding SIM profiles to software trust anchors (e.g., TEEs, confidential VMs)
\end{itemize}

To enable the model proposed in this paper, GSMA and 3GPP would need to consider new profile types, API-level access controls, and attestation-based access models.

\subsection{Comparison with Other Identity Mechanisms}

The identity ecosystem in cloud and enterprise environments is dominated by application-layer technologies such as API keys, OAuth2 tokens, mutual TLS certificates, and hardware TPM-backed keys. While these models are suitable for web services and human users, they have key limitations when applied to autonomous agents operating in sensitive, distributed environments.

\textbf{API Keys and OAuth2 Tokens.} These are bearer tokens: anyone in possession of the token can act as the agent. They can be copied, stolen, or injected into other processes. Their security depends entirely on application logic, not on infrastructure-level enforcement.

\textbf{X.509 Certificates.} Certificates tied to local private keys offer stronger identity assurance than bearer tokens, but the security of the private key still depends on local key storage (e.g., file system, software vault). This makes them vulnerable to compromise, especially in containerized or multi-tenant cloud environments.

\textbf{TPMs and Attestation.} Trusted Platform Modules (TPMs) and secure enclaves can securely store private keys and perform attestable operations. However, their use is limited to environments that directly own the TPM. There is no global, cross-enterprise trust model for TPM identity similar to the GSMA root-of-trust used in SIM ecosystems.

\textbf{SIM-Based Identity.} In contrast, the architecture proposed here offers:
\begin{itemize}
    \item Infrastructure-level enforcement: all operations occur in a Telco-controlled secure element.
    \item Non-exportability: the key is never exposed, even to the agent runtime.
    \item Lifecycle control: SIM profiles can be provisioned, revoked, and monitored using mature, telecom-grade protocols.
    \item Global trust: existing GSMA root-of-trust infrastructure enables federated use across networks and operators.
\end{itemize}

In short, this model brings telecom-grade identity assurance—originally built for billions of mobile devices—to a new class of digital entities: autonomous AI agents.

\section{Conclusion}

As AI agents become first-class actors in enterprise, industrial, and financial systems, the need for secure, infrastructure-grade identity becomes increasingly critical.Existing identity frameworks designed for human users, physical devices, or application-level access control are poorly suited for dynamic, autonomous agents operating in distributed, cloud-native environments.

In this paper, we proposed a novel architectural model that extends telecom-grade SIM identity to software-based agents by decoupling the SIM credential from physical hardware and hosting it within Telco-controlled secure infrastructure. This model leverages existing standards for eSIM provisioning, SIM-based authentication, and remote attestation to provide agents with strong, auditable, non-exportable identities that can be policy-controlled and revoked in real time.

We demonstrated how this identity-as-a-service architecture enables scalable authentication, compliance-grade auditing, and trustable inter-agent communication in scenarios such as enterprise automation, financial services, and industrial edge AI. We also identified specific gaps in GSMA and 3GPP standards, and outlined the necessary technical extensions to support software-native agent identity within the existing telecom trust fabric.

Looking forward, we believe this architecture represents a foundation for agent-scale identity systems. Future work may explore agent-to-agent trust protocols, agent identity federation across Telcos, and standardization within GSMA or IETF working groups. As telecom infrastructure evolves to serve not just devices, but intelligent software agents, identity will become both a product and a platform.

\end{document}